\documentstyle[12pt]{article}
\def\beq{\begin{equation}}
\def\eeq{\end{equation}}
\def\bea{\begin{eqnarray}}
\def\eea{\end{eqnarray}}

\def\ba{\begin{array}}
\def\ea{\end{array}}
\def\ds{\displaystyle}

\def\,{\"{U}}
\def\6{\.{I}}

\begin{document}
\baselineskip 0.7cm

\title{ {Exact \Large{ Supersymmetric Solution of Schr\"{o}dinger Equation
for central confining Potentials by using the Nikiforov-Uvarov
Method }}}

\author{Metin Akta\c{s} and Ramazan Sever$\thanks{Corresponding author:
sever@metu.edu.tr}$\\
Department of Physics, Middle East Technical University \\
06531 Ankara, Turkey }

\date{\today}

\maketitle

\begin{abstract}

We present the exact supersymmetric solution of Schr\"{o}dinger
equation with the Morse, P\"{o}schl-Teller and Hulth\'{e}n
potentials by using the Nikiforov-Uvarov method. Eigenfunctions
and corresponding energy eigenvalues are calculated for the first
six excited states. Results are
in good agreement with the ones obtained before.\\

\end{abstract}

\newpage
\section{Introduction}
\noindent Supersymmetric quantum mechanics (SUSYQM) is a framework
used to determine energy eigenvalues and eigenfunctions of quantum
mechanical problems. Various methods have been used in their
solutions [1]. One of them is the factorization method introduced by
Schr\"{o}dinger to study the hydrogen atom problem [2]. This method was
developed later by Infeld and Hull to categorize the analytically
solvable potential problems [3]. Also it was used to derive in the
solutions of certain non-linear equations [4]. The others are
approximation methods
known as $1/N$ expansion [5], $\delta-expansion$ [6], supersymmetric WKB
(SWKB) [7, 8] and
variational methods [9]. About two decades ago,
Gendenshtein introduced a concept ``shape invariance''
to obtain the exact energy eigenvalues of the Schr\"{o}dinger
equation [12, 13]. It provides us a simple approach to the solution when
the potentials have shape invariance property. Recently, an
alternative method known as the Nikiforov-Uvarov method
(NU-method) has been introduced for solving the Schr\"{o}dinger
equation (SE). There have been several applications of SE with some
well-known potentials [24, 25, 26], Dirac and
Klein-Gordon equations for a Coulomb potential by using this method as
well [27].

This article is organised as follows: In Sec. (2), we give a
summary of SUSYQM. In Sec. (3), we introduce the NU-method to
solve SE. In Sec. (4), solution of Morse, P\"{o}schl-Teller and
Hulth\'{e}n potentials are obtained. Numerical results are given
in the tables. Finally, we discuss the results in the conclusion.

\section{Summary of SUSYQM}
\noindent Supersymmetric algebra allows us to write Hamiltonians
as [1]

\begin{equation}
H_{\pm}=-\frac{\hbar^{2}}{2m}\frac{d^{2}}{dx^{2}}+
V_{\pm}(x),
\end{equation}

\noindent where
The supersymmetric partner potentials $V_{\pm}(x)$
in terms of the superpotential $W(x)$ are given by

\begin{equation}
V_{\pm}(x)=W^{2}\pm\frac{\hbar}{\sqrt{2m}}
\frac{dW}{dx}.
\end{equation}

\noindent
The superpotential has a definition

\begin{equation}
W(x)=-\frac{\hbar}{\sqrt{2m}}\left[\frac{d\ln\Psi_{0}^{(0)}(x)}{dx}\right],
\end{equation}

\noindent
where, $\Psi_{0}^{(0)}(x)$ denotes the ground state wave
function that satisfies the relation

\beq
\Psi_{0}^{(0)}(x)=N_{0}~\exp\left[-\frac{\sqrt{2m}}{\hbar}
\int^{x} W(x^{\prime})~dx^{\prime}\right].
\eeq

\noindent
The Hamiltonian $H_{\pm}$ can also be written in terms of the
bosonic operators $A^{-}$ and $A^{+}$

\begin{equation}
H_{\pm}=A^{\mp}~A^{\pm},
\end{equation}

\noindent
where

\begin{equation}
A^{\pm}=\pm\frac{\hbar}{\sqrt{2m}}\frac{d}{dx}+W(x).
\end{equation}

It is remarkable result that the energy eigenvalues of $H_{-}$
and $H_{+}$ are identical except for the ground state. In the
case of unbroken supersymmetry, the ground state energy of the
Hamiltonian $H_{-}$ is zero $\left(E_{0}^{(0)}=0\right)$ [14].
In the factorization of the Hamiltonian, the Eqs. (1), (5)
and (6) are used respectively. Hence, we obtain

\begin{eqnarray}
H_{1}(x)&=&-\frac{{\hbar}^{2}}{2m}\frac{d^{2}}{dx^{2}}+V_{1}(x)\nonumber\\[0.2cm]
        &=&(A_{1}^{+}~A_{1}^{-})+E_{1}^{(0)}.
\end{eqnarray}

\smallskip
\noindent
Comparing each side of the Eq. (7) term by term, we get
the Riccati equation for the superpotential $W_{1}\left(x\right)$

\begin{equation}
W_{1}^{2}-W_{1}^{'}=\frac{2m}{{\hbar}^{2}}\left(V_{1}(x)-E_{1}^{(0)}\right).
\end{equation}

\smallskip
Let us now construct the supersymmetric partner
Hamiltonian $H_{2}$ as

\begin{eqnarray}
H_{2}(x)&=&-\frac{{\hbar}^{2}}{2m}\frac{d^{2}}{dx^{2}}+V_{2}(x)\nonumber\\[0.2cm]
        &=&\left(A_{2}^{-}~A_{2}^{+}\right)+E_{2}^{(0)},
\end{eqnarray}

\noindent
and Riccati equation takes
\begin{equation}
W_{2}^{2}+W_{2}^{'}=\frac{2m}{{\hbar}^{2}}\left(V_{2}(x)-E_{2}^{(0)}\right).
\end{equation}

\smallskip
\noindent Similarly, one can write in general the  Riccati
equation and Hamiltonians by iteration as

\smallskip
\begin{eqnarray}
W_{n}^{2}\pm
W_{n}^{'}&=&\frac{2m}{{\hbar}^{2}}\left(V_{n}(x)-E_{n}^{(0)}\right)\nonumber\\[0.2cm]
         &=&\left(A_{n}^{\pm}~A_{n}^{\mp}\right)+E_{n}^{(0)},
\end{eqnarray}

\noindent
and

\begin{eqnarray}
H_{n}(x)&=&-\frac{{\hbar}^{2}}{2m}\frac{d^{2}}{dx^{2}}+V_{n}(x)\nonumber\\[0.2cm]
        &=& A_{n}^{+}~A_{n}^{-}+E_{n}^{(0)},\quad\quad n=1,2,3,\ldots
\end{eqnarray}

\noindent where \begin{equation}
A_{n}^{\pm}=\pm\frac{\hbar}{\sqrt{2m}}\frac{d}{dx}+\frac{d}
{dx}\left(\ln\Psi_{n}^{(0)}(x)\right).
\end{equation}

\smallskip
\noindent
Because of the SUSY unbroken case, the partner Hamiltonians
satisfy the following expressions [9, 14]

\begin{equation}
E_{n+1}^{(0)}=E_{n}^{(1)}, \quad with \quad E_{0}^{(0)}=0,\quad
n=0,1,2,\ldots
\end{equation}

\smallskip
\noindent and also the wave function with the same eigenvalue can
be written as [14]

\begin{equation}
\Psi_{n}^{(1)}=\frac{A^{-}~\Psi_{n+1}^{(0)}}{\sqrt {E_{n}^{(0)}}},
\end{equation}

\noindent
with

\begin{equation}
\Psi_{n+1}^{(0)}=\frac{A^{+}~\Psi_{n}^{(1)}}{\sqrt {E_{n}^{(0)}}}.
\end{equation}

\smallskip
This procedure is known as the hierarchy of Hamiltonians.

\section{The Nikiforov-Uvarov Method}
\noindent The NU-method reduces the second order differential equations
(ODEs) to the hypergeometric type with an appropriate coordinate
transformation $x=x(s)$ as

\begin{equation}
\Psi^{\prime\prime}(s)+\frac{\tilde{\tau}(s)}{\sigma(s)}
\Psi^{\prime}(s)+\frac{\tilde{\sigma}(s)}{{\sigma}^{2}(s)}\Psi(s)=0
\end{equation}

\noindent
where ${\sigma}(s)$ and ${\tilde{\sigma}(s)}$ are
polynomials with at most second degree, and
${\tilde{\tau}(s)}$ is a polynomial with at most first
degree [24, 25, 26, 27]. If we
take the following factorization

\begin{equation}
\Psi(s)=\phi(s)~y(s),
\end{equation}

\noindent
the Eq. (17) becomes [27]

\begin{equation}
\sigma(s)~y^{\prime\prime}(s)+\tau(s)~y^{\prime}(s)+\Lambda~y(s)=0,
\end{equation}

\noindent
where

\begin{equation}
\sigma(s)=\pi(s)\frac{d}{ds}(\ln\phi(s)),
\end{equation}
\noindent

\noindent
and
\begin{equation}
\tau(s)=\tilde{\tau}(s)+2\pi(s).
\end{equation}

\noindent Also, $\Lambda$ is defined as

\begin{equation}
\Lambda_{n}+n\tau^{\prime}+\frac{\left[n(n-1)\sigma^{\prime\prime}\right]}{2}=0,
\quad n=0,1,2,\ldots
\end{equation}

\smallskip
The energy eigenvalues can be calculated from the above equation. We
first have to determine $\pi(s)$ and $\Lambda$ by defining

\begin{equation}
k=\Lambda-\pi^{\prime}(s).
\end{equation}

\noindent
Solving the quadratic equation for $\pi(s)$ with the
Eq. (23), we get

\smallskip
\begin{equation}
\pi(s)=\left(\frac{\sigma^{\prime}-\tilde{\tau}}{2}\right)\pm
\sqrt{\left(\frac{\sigma^{\prime}-\tilde{\tau}}{2}\right)^{2}-
\tilde{\sigma}+k\sigma}.
\end{equation}

\noindent Here, $\pi(s)$ is a polynomial with the parameter $s$
and prime factors denote the differentials at first degree. The
determination of $k$ is the essential point in the calculation of
$\pi(s)$. It is simply defined as by setting the discriminant of
the square root must be zero [27]. Therefore, we obtain a general
quadratic equation for $k$.

\smallskip
The determination of the wave function is now in order. We consider
the Eq. (20) and the Rodrigues relation

\begin{equation}
y_{n}(s)=\frac{C_{n}}{\rho(s)}\frac{d^{n}}{ds^{n}}
\left[\sigma^{n}(s)\rho(s)\right],
\end{equation}

\smallskip
\noindent where $C_{n}$ is normalizable constant and the weight
function $\rho(s)$ satisfy the following relation

\begin{equation}
\frac{d}{ds}\left[\sigma(s)~\rho(s)\right]=\tau(s)~\rho(s).
\end{equation}
\smallskip
\noindent
The Eq. (25) refers to the classical orthogonal
polynomials that have many important properties especially
orthogonality relation can be defined as

\smallskip
\begin{equation}
\int_{a}^{b}y_{n}(s)~ y_{m}(s)~\rho(s)~ds=0,\quad m\neq n.
\end{equation}

\section{Calculations}
\noindent

We will use the Nikiforov-Uvarov method by getting the hypergeometric or
confluent hypergeometric form of the  Schr\"{o}dinger equation with the
following potentials for $s-states$ only.

\subsection{Morse Potential}

\noindent
The Morse potential is

\begin{equation}
V_{M}(r)=D\left(e^{-2~a~ x}-2e^{-a~x}\right),
\end{equation}

\noindent
with ${\ds x={(r-r_{0})}/{r_{0}}}$ [23]. Here, D denotes the dissociation
energy parameter and $r_{0}$ is the equilibrium distance between nucleus.
Its supersymmetric form becomes [1]

\begin{equation}
W(x)=A-Be^{-a~x}.
\end{equation}

\noindent
Thus, we first get the superpartner potentials as

\begin{equation}
V_{\pm}(x,A,B)=A^{2}+B^{2}e^{-2~a~x}-
2B\left(A\mp\frac{\hbar~a}{\sqrt{8m}}\right)e^{-a~x}.
\end{equation}

\smallskip
\noindent
From the Eq. (1), Schr\"{o}dinger equation can be written as [8]

\begin{equation}
H_{\pm}\Psi=E_{\pm}\Psi
\end{equation}

\noindent
or explicitly

\begin{equation}
\frac{d^{2}\Psi}{dx^{2}}+\frac{2m}{\hbar^{2}}\left[E_{-}-V_{-}\right]\Psi=0.
\end{equation}

\smallskip
\noindent
Using the Eq. (30) for $V_{-}$, we get

\begin{equation}
\Psi^{\prime\prime}(x)+{\frac{2m}{\hbar^{2}}}\left[\bar E-\bar Be^{-2~a~x}+
\bar Ce^{-a~x}\right]\Psi(x)=0,
\end{equation}

\smallskip
\noindent where $\bar E=E_{-}-A^{2}$, $\bar B=B^{2}$ and ${\ds
\bar C=2B\left(A+\frac{\hbar~a}{\sqrt{8m}}\right)}$. By
introducing a transformation

\smallskip
\begin{equation}
-s=e^{-a~x},
\end{equation}

\noindent
the Eq. (33) takes the form

\begin{equation}
\Psi^{\prime\prime}(s)+\frac{1}{s}\Psi^{\prime}(s)
+\frac{1}{s^{2}}\left[\beta^{2}-\gamma^{2}s-\alpha^{2}s^{2}\right]\Psi(s)=0,
\end{equation}

\noindent
where

\begin{equation}
\alpha^{2}=\frac{2m\bar B}{a^{2}\hbar^{2}}
 ,\quad\beta^{2}=\frac{2m\bar E}{a^{2}\hbar^{2}}
\left(E^{'}<0\right)\quad and \quad
\gamma^{2}=\frac{2m\bar C}{a^{2}\hbar^{2}}.
\end{equation}

\smallskip
\noindent
Comparing the Eq. (35) with the Eq. (17), we obtain

\smallskip
\begin{equation}
\sigma(s)=s,\quad\tilde{\tau}(s)=1 \quad and\quad\tilde{\sigma}(s)=
\left(-\beta^{2}-\gamma^{2}s-\alpha^{2}s^{2}\right).
\end{equation}

\smallskip
\noindent
Substituting these polynomials into the Eq. (24), we get

\smallskip
\begin{equation}
\pi(s)=\pm\sqrt{\alpha^{2}s^{2}+(k+\gamma^{2})s+\beta^{2}}.
\end{equation}

\smallskip
\noindent The constant $k$ is determined as

\begin{equation}
k_{1,~2}=-\gamma^{2}\pm 2\alpha\beta,
\end{equation}

\smallskip
\noindent and we have

\begin{eqnarray}
\pi(s)=
\left\{
\begin{array}{ll}
\pm (\alpha s-\beta), ~~~&\mbox{$for ~~~ k=-\gamma^2-2\alpha\beta$}\\[0.5cm]
\pm (\alpha s+\beta), ~~~&\mbox{$for ~~~ k=-\gamma ^{2}+2\alpha\beta$}.
\end{array}
\right.
\end{eqnarray}

\noindent
A proper value for $\pi(s)$ is chosen, so that the function

\begin{equation}
\tau(s)=(1+2\beta)-2\alpha s,
\end{equation}

\noindent
has a negative derivative [27]. By using the Eq. (22), we can find

\begin{eqnarray}
  \Lambda&=&-\gamma^{2}-2\alpha\beta-\alpha \nonumber\\
         &=&2\alpha n.
\end{eqnarray}

\noindent
Thus, we simply get the energy eigenvalues as

\smallskip
\begin{equation}
\bar
E_{n,~\ell=0}=\frac{a^{2}\hbar^{2}}{2m}\left[\bar
D-(n+\frac{1}{2})\right]^{2},
\end{equation}

\smallskip
\noindent
where
${\ds \bar D=\gamma^{2}/2\alpha}$. By setting $\alpha=1$, this
equation reduces to the Eq. (10) as [28] for $s-states$. The Eq.
(43) can be seperable. Its square term refers to the anharmonic oscillator
correction and the other one corresponds to the harmonic oscillator
solution.

\smallskip
Now, we are going to determine the eigenfunctions for this potential. By
considering the Eq. (18) and using the Eq. (20), we obtain

\begin{equation}
\phi(s)=s^{\beta}~e^{-\alpha s}.
\end{equation}

\noindent
By using the Eqs. (26) and (25), we obtain

\smallskip
\begin{equation}
y_{n}(s)=\frac{C_{n}}{\rho(s)}\frac{d^{n}}{ds^{n}}\left[s^{n}\rho(s)\right],
\end{equation}

\smallskip
\noindent
where ${\ds\rho(s)=s^{2\beta}~e^{-2\alpha s}}$. The Eq.
(45) stands for the associated Laguerre polynomials, that is

\begin{equation}
y_{n}(s)\equiv L_{n}^{\mathit t}(s),
\end{equation}

\smallskip
\noindent where $t=2\beta$. Hence, we can write the wave
function in the final form

\smallskip
\begin{equation}
\Psi_{n}(x)=C_{n}~s^{\beta}~e^{-\eta s}~L_{n}^{\mathit t}(s),
\end{equation}

\smallskip
\noindent with ${\ds s=-e^{a~x}}$. It is normalizable. Using
the Eq. (27), the normalization constant can be found as

\smallskip
\begin{equation}
C_{n}=\sqrt{\frac{n!}{(n+\beta+\frac{1}{2})(n+2\beta)!}}, \quad
n=0,1,2\ldots
\end{equation}

\subsection{P\"{o}schl-Teller Potential}

\noindent
The P\"{o}schl-Teller potential is

\begin{equation}
V_{PT}(x)=-\frac{U_{0}}{cosh^{2}\alpha x},
\end{equation}

\noindent
where $U_{0}=\lambda~(\lambda-1)>0$ [8]. Also, its superpotential
potential is [1]

\smallskip
\begin{equation}
W(x)=A\tanh\alpha x.
\end{equation}

\smallskip
\noindent
From Eq. (2), we get its superpartners

\smallskip
\begin{equation}
V_{\pm}(x,A)=A^{2}-A\left(A\mp\frac{{\alpha}{\hbar}}{\sqrt{2m}}\right)
\frac{1}{cosh^{2}\alpha x}.
\end{equation}

\smallskip
\noindent
Thus, we can write the Schr\"{o}dinger equation as

\smallskip
\begin{equation}
\Psi^{\prime\prime}(x)+\frac{2m}{\hbar^{2}}\left[\widetilde E-\frac{\kappa}
{cosh^{2}\alpha x}\right]\Psi(x)=0,
\end{equation}

\bigskip
\noindent
where $\widetilde E=E_{-}-A^{2}$ and ${\ds
\kappa=A\left(A+\frac{{\alpha}{\hbar}}
{\sqrt{2m}}\right)}$. Introducing a transformation

\begin{equation}
s=\tanh\alpha x,
\end{equation}

\noindent
we rewrite the Eq. (52)

\begin{equation}
\Psi^{\prime\prime}(s)+\frac{-2s}{(1-s^{2})}\Psi^{\prime}(s)+
\frac{1}{(1-s^{2})^{2}}
\left[-\beta^{2}+\gamma^{2}(1-s^{2})\right]\Psi(s)=0,
\end{equation}

\noindent
where

\begin{equation}
\beta^{2}=\frac{2m\widetilde E}{\alpha^{2}\hbar^{2}}~(\widetilde E<0)\quad
and \quad\gamma^{2}=\frac{2m\kappa}{\alpha^{2}\hbar^{2}}.
\end{equation}

\smallskip
\noindent By comparing the Eq. (54) with the Eq. (17), we
determine polynomials as

\smallskip
\begin{equation}
\sigma(s)=(1-s^{2}),\quad\tilde{\tau}(s)=-2s
\quad and\quad\tilde{\sigma}(s)=
-\beta^{2}+\gamma^{2}(1-s^{2}).
\end{equation}

\smallskip
\noindent Substituting them into the Eq. (24), we obtain

\begin{equation}
\pi(s)=\pm\sqrt{\beta^{2}-\gamma^{2}(1-s^{2})+k(1-s^{2})}.
\end{equation}

\noindent The constant $k$ is determined in the same way.
Therefore, we get

\begin{eqnarray}
\pi(s)=
\left\{\begin{array}{ll}
\pm\beta, ~~~~~&\mbox{$for ~~~~ k=\gamma^{2}$}\\[0.5cm]
\pm\beta s, ~~~&\mbox{$for ~~~~ k=\gamma^{2}-\beta^{2}$}.
\end{array}
\right.
\end{eqnarray}

\noindent Here we choose the proper value, so that
\begin{equation}
\tau(s)=-2(1+\beta)s,
\end{equation}

\noindent
has a negative derivative. From the Eq. (22), we calculate

\begin{eqnarray}
\Lambda&=&\gamma^{2}-\beta^{2}-\beta \nonumber\\
       &=&n^{2}+n+2n\beta.
\end{eqnarray}

\smallskip
\noindent
Hence, the energy eigenvalues are found as

\smallskip
\begin{equation}
\widetilde E_{n}=A^{2}-\frac{\hbar^{2}\alpha^{2}}{2m}\left[-(n+\frac{1}{2})
+\frac{1}{2}\sqrt{1+4\gamma^{2}}\right]^{2}.
\end{equation}

\smallskip
The wave function $\Psi_{n}(x)$ is obtained from the Eq. (18) by
taking~$\pi(s)=-\beta s$~ as follows. We first get

\begin{equation}
\phi(s)=(1-s^{2})^{{\beta}/{2}},
\end{equation}

\smallskip
\noindent and using the Eqs. (26) and (25), we find

\begin{equation}
y_{n}(s)=\frac{C_{n}}{\rho(s)}\frac{d^{n}}{ds^{n}}\left[(1-s^{2})^{n}
\rho(s)\right],
\end{equation}

\noindent
where ${\ds\rho(s)=(1-s^{2})^{\beta}}$. The Eq. (63) stands for
the Jacobi polynomials as

\begin{equation}
y_{n}\equiv P_{n}^{(\beta,~\beta)}(s).
\end{equation}

\noindent
Hence, $\Psi_{n}(x)$ can be written in the following form

\begin{equation}
\Psi_{n}(x)=C_{n}~(1-s^{2})^{{\beta}/{2}}~P_{n}^{(\beta,~\beta)}(s),
\end{equation}

\smallskip
\noindent with $s=\tanh\alpha~x$. Considering the Eq. (27), the
normalization constant is obtained as

\smallskip
\begin{equation}
C_{n}=\frac{1}{{2^{\beta}}~(n+\beta)!}\sqrt{\frac{(2n+2\beta+1)}{2}
n!(2n+\beta)!}.
\end{equation}

\noindent
where $n,~\beta\ge0$.

\subsection{Hulth\'{e}n Potential}
\noindent This potential can be solved exactly for $s-states$
only. This is due to the similarity between Coulomb and
Hulth\'{e}n potentials. It plays an important role in the
applications of quantum scattering theory. The Hulth\'{e}n
potential is given by [23]

\begin{equation}
V_{0}^{H}=-V_{0}\frac{e^{-\lambda~x}}{(1-e^{-\delta~x})},
\end{equation}

\noindent where $\delta=1/a$, is the screening parameter. We get the
supersymmetric form the potential for
$s-states$ [10]

\begin{equation}
W_{1}=\bar a+ \bar b\frac{e^{-\delta~x}}{(1-e^{-\delta~x})}.
\end{equation}

\noindent
Here, $\bar a$ and $\bar b$ are arbitrary constants. We can also write the
supersymmetric partner of the potential as [11]

\begin{equation}
V_{1}^{H}=V_{0}^{H}+\frac {V_{0}^{2}~e^{-\delta~x}}
{(1-e^{-\delta~x})^{2}}.
\end{equation}

\smallskip
\noindent
The second term in Eq. (69) behaves like
centrifugal barrier [23]. The SE has the form

\begin{equation}
\Psi^{\prime\prime}(x)+\frac{2m}{\hbar^{2}}\left[\overline {E}
-V_{1}^{H}(x)\right]\Psi(x)=0.
\end{equation}

\noindent
Using the transformation

\begin{equation}
s=e^{-\delta~x},
\end{equation}

\noindent
we rewrite

\begin{equation}
\Psi^{\prime\prime}(s)+\frac{(1-s)}{\left[s(1-s)\right]}\Psi^{\prime}(s)+
\frac{1}{\left[s(1-s)\right]^{2}}\left[-(\varepsilon^{2}+\beta^{2})s^{2}+
(2\varepsilon^{2}+\beta^{2}-\gamma^{2})s-\varepsilon^{2}\right]\Psi(s)=0,
\end{equation}

\noindent where
\begin{equation}
\varepsilon^{2}=\frac{2m\overline{E}}{\delta^{2}\hbar^{2}}(\overline
{E}>0) ,\quad \beta^{2}=\frac{2mV_{0}}{\delta^{2}\hbar^{2}}
\quad and \quad\gamma^{2}=\frac{2mV_{0}^{2}}{\delta^{2}\hbar^{2}}.
\end{equation}

\smallskip
\noindent
By comparing the Eq. (72) with the Eq. (17), we get

\smallskip
\begin{equation}
\sigma(s)=s(1-s),\quad\tilde{\tau}(s)=1-s\quad
and\quad\tilde{\sigma}(s)=-(\varepsilon^{2}+\beta^{2})s^{2}+
(2\varepsilon^{2}+\beta^{2}-\gamma^{2})s-\varepsilon^{2}.
\end{equation}

\noindent
Substituting them into the Eq. (24), we obtain

\begin{equation}
\pi(s)=-\frac{1}{2}
s\pm\frac{1}{2}\sqrt{4\left(\varepsilon^{2}+\beta^{2}-k+
\frac{1}{4}\right) s^{2}-4\left(2\varepsilon^{2}+\beta^{2}
-\gamma^{2}-k\right)s+4\varepsilon^{2}}.
\end{equation}

\smallskip
\noindent
From Eq. (75) $k$ is determined as

\begin{equation}
k_{1,~2}=-\gamma^{2}+\beta^{2}\pm\varepsilon\sqrt{1+4\gamma^{2}}.
\end{equation}

\smallskip
\noindent
Following the same procedure, we get

\begin{equation}
\pi(s)=-\frac{1}{2} s\pm\frac{1}{2}\left[\left(2\varepsilon
+\sqrt{1+4\gamma^{2}}\right) s-2\varepsilon\right],
\end{equation}

\smallskip
\noindent and the energy eigenvalues for the supersymmetric
Hulth\'{e}n potential becomes

\smallskip
\begin{equation}
\overline {E}_{\bar
n,~\ell=0}=-V_{0}\left[\frac{\beta^{2}-\bar n^{2}}{2
~\bar n~\beta}\right]^{2},\quad \bar n=1,2,\ldots
\end{equation}

\bigskip
\noindent Here, $\beta^{2}=2V_{0}/\delta^{2}$ with $(\hbar=m=1)$ and
${\ds \bar
n=\left[(n+\frac{1}{2})-\frac{1}{2}\sqrt{1+4\gamma^{2}}\right]}$.
If the limit $\gamma\rightarrow{0}$ is chosen, the energy eigenvalue
reduces to the form obtained from the usual solution of the Hulth\'{e}n
potential.

\smallskip
The wave functions can now be obtained similarly from the
Eq. (18). Using the Eq. (20), we have

\smallskip
\begin{equation}
\phi(s)=s^{\varepsilon}~(1-s)^{{\mu}/{2}},
\end{equation}

\noindent
where $\mu=1+\sqrt{1+4\gamma^{2}}$. The Eqs.
(26) and (25) lead to

\begin{equation}
y_{n}(s)=\frac{C_{n}}{\rho(s)}\frac{d^{n}}{ds^{n}}\left[s^{n}(1-s)^{n}
\rho(s)\right].
\end{equation}

\smallskip
\noindent
Here, ${\ds\rho(s)=s^{2\varepsilon}~(1-s)^{\mu-1}}$. It stands for the
Jacobi polynomials as [25]

\begin{equation}
y_{n}(s)\simeq P_{n}^{(2\varepsilon,~\mu-1)}(1-2s).
\end{equation}

\smallskip
\noindent
Thus, the final form of the wave function can also be
written in terms of the Jacobi polynomials resulting

\begin{equation}
\Psi_{n}(x)=C_{n}~s^{\varepsilon}~(1-s)^{{\mu}/{2}}
~P_{n}^{(2\varepsilon,~\mu-1)}(1-2s),
\end{equation}

\smallskip
\noindent with ${\ds s=e^{-\delta~x}}$, and also the normalization
constant $C_{n}$.

\section{Conclusions}
\noindent We have obtained the exact supersymmetric solution of
some central confining potentials by  applying the
Nikiforov-Uvarov Method. The eigenfunctions and corresponding
energy egenvalues of the these three well-known shape invariant
potentials, i.e. Morse, P\"{o}schl-Teller and Hulth\'{e}n are
calculated analytically. All the wave functions are physical. We
present numerical results in tabular form for $\ell=0$. In Table
I, we list energy eigenvalues of $H_{2}$ molecule by taking
$D=4.7446~eV$, $a=1.9425~\AA^{-1}$ and $m=0.50391~amu$ for ground
state and first eleven excited states. The energy difference
between the successive states decreases. In Table II, the six
excited energy states of the the potential are given for various
values of $n$ and $\lambda$ with
$A=(1+\sqrt{1+4~\lambda~(\lambda-1)})/2$. In Table III, the first
five excited energy levels are tabulated for $s-states$ with
different values of screening parameter $\delta$. Our results are
in good agreement with the ones obtained by the other methods.

\newpage

{\bf Table I:}~~{\small Vibrational (non-rotating) energy
eigenvalues of
the Morse potential for $H_{2}$ molecule.}\\

\begin{tabular}{cccc}\hline\hline
\\ $\mathbf{Quantum~no~(n)}
\hspace*{0.4cm}
$ & $
\hspace*{0.4cm}
\mathbf{Eigenvalues}~[28]
\hspace*{0.4cm}
$ & $
\hspace*{0.4cm}
\mathbf{\bar E_{n}}~(Our~work)
\hspace*{0.4cm}
$ & $
\hspace*{0.4cm}
\mathbf{\bar E_{n+1}-\bar E_{n}}$ \\[0.3cm]\hline
0~~&~~-4.47600~~&~~-4.47610~~&~~0.51354 \\[0.2cm]\hline
1~~&~~~~&~~-3.96256~~&~~0.48226 \\[0.2cm]
2~~&~~~~&~~-3.48030~~&~~0.45100 \\[0.2cm]
3~~&~~~~&~~-3.02930~~&~~0.41968 \\[0.2cm]
4~~&~~~~&~~-2.60962~~&~~0.38842 \\[0.2cm]\hline
5~~&~~-2.22050~~&~~-2.22120~~&~~0.35713 \\[0.2cm]
6~~&~~-1.86330~~&~~-1.86407~~&~~0.32587 \\[0.2cm]
7~~&~~-1.53740~~&~~-1.53820~~&~~0.29456 \\[0.2cm]\hline
8~~&~~~~&~~-1.24364~~&~~0.26330 \\[0.2cm]
9~~&~~~~&~~-0.98034~~&~~0.23201 \\[0.2cm]
10~~&~~~~&~~-0.74833~~&~~0.20073 \\[0.2cm]
11~~&~~~&~~-0.54760~~&~~~ \\[0.2cm]\hline
\end{tabular}

\newpage

{\bf Table II:}~~{\small Eigenvalues of the  P\"{o}schl-Teller potential
with$A=\left(1+\sqrt{1+4~\lambda~(\lambda-1)}\right)/2$ setting by
$(\hbar=\alpha=2m=1, A^{2}=0~and~ \kappa=\gamma^{2})$~in~Eq.~(61)~[29].}\\

\begin{tabular}{|lcccccr}\hline\hline
$\mbox{\boldmath$\lambda$}
\hspace*{1cm}
\put(9,-8){\line(-3,5){14}}
$ & $
-\mathbf{\widetilde E_{1}}
\hspace*{0.5cm}
$ & $
\hspace*{0.5cm}
-\mathbf{\widetilde E_{2}}
\hspace*{0.5cm}
$ & $
\hspace*{0.5cm}
-\mathbf{\widetilde E_{3}}
\hspace*{0.5cm}
$ & $
\hspace*{0.5cm}
-\mathbf{\widetilde E_{4}}
\hspace*{0.5cm}
$ & $
\hspace*{0.5cm}
-\mathbf{\widetilde E_{5}}
\hspace*{0.5cm}
$ & $
\hspace*{0.5cm}
-\mathbf{\widetilde E_{6}} $\\[0.5cm]\hline
\\
$\mbox{\boldmath$2.0$}$~~&~~0.0~~&~~1.0~~&~~4.0~~&~~9.0~~&~~16.0~~&~~25.0\\[0.2cm]
\\
$\mbox{\boldmath$6.0$}$~~&~~&~~0.0~~&~~1.0~~&~~4.0~~&~~9.0~~&~~16.0
\\[0.2cm]
\\
$\mbox{\boldmath$12.0$}$~~&~~&~~&~~0.0~~&~~1.0~~&~~4.0~~&~~9.0 \\[0.2cm]
\\
$\mbox{\boldmath$20.0$}$~~&~~&~~&~~&~~0.0~~&~~1.0~~&~~4.0 \\[0.2cm]
\\
$\mbox{\boldmath$30.0$}$~~&~~&~~&~~&~~&~~0.0~~&~~1.0 \\[0.2cm]
\\
$\mbox{\boldmath$42.0$}$~~&~~&~~&~~&~~&~~&~~0.0 \\[0.2cm]\hline
\end{tabular}

\newpage
{\bf Table III:}~~{\small Eigenvalues of the Hulth\'{e}n potential for
several values of screening parameter $\delta$, substituting
$V_{0}=\delta/4$~in~$\beta^{2}$~of~the~Eq.~(78).}\\[0.1cm]

\begin{tabular}{cccc}\hline\hline
\\ $\mathbf{Quantum~no~(n)}
\hspace*{0.5cm}
$ & $
\hspace*{0.5cm}
-\mathbf{E_{n}~[30]}
\hspace*{0.5cm}
$ & $
\hspace*{0.5cm}
-\mathbf{E_{exact}}
\hspace*{0.5cm}
$ & $
\hspace*{0.5cm}
-\mathbf{\overline {E}_{n}~(Our~work)}$ \\[0.1cm]\hline
\\
\put(65, 0){$\mbox{\boldmath$\delta=0.002$}$}\\[0.1cm]\hline
1~~&~~0.4990005~~&~~0.4990005~~&~~0.4990005 \\[0.05cm]
2~~&~~0.1240020~~&~~0.1240020~~&~~0.1240020 \\[0.05cm]
3~~&~~0.0545601~~&~~0.0545601~~&~~0.0545601 \\[0.05cm]
4~~&~~0.0302580~~&~~0.0302580~~&~~0.0302580 \\[0.05cm]
5~~&~~~~&~~&~~0.0012500 \\[0.1cm]\hline
\\
\put(65, 0){$\mbox{\boldmath$\delta=0.01$}$}\\[0.1cm]\hline
1~~&~~0.4950125~~&~~0.4950125~~&~~0.4950125 \\[0.05cm]
2~~&~~0.1200500~~&~~0.1200500~~&~~0.1200500 \\[0.05cm]
3~~&~~0.0506681~~&~~0.0506681~~&~~0.0506681 \\[0.05cm]
4~~&~~0.0264501~~&~~0.0264500~~&~~0.0264500 \\[0.05cm]
5~~&~~0.0153128~~&~~0.0153125~~&~~0.0153125 \\[0.05cm]\hline
\\
\put(65, 0){$\mbox{\boldmath$\delta=0.05$}$}\\[0.1cm]\hline
1~~&~~0.4753125~~&~~0.4753125~~&~~0.4753125 \\[0.05cm]
2~~&~~0.1012503~~&~~0.1012500~~&~~0.1012500 \\[0.05cm]
3~~&~~0.0333746~~&~~0.0333681~~&~~0.0333681 \\[0.05cm]
4~~&~~0.0113035~~&~~0.0112500~~&~~0.0112500 \\[0.05cm]
5~~&~~~~&~~~~~&~~0.0028125 \\[0.1cm]\hline
\\
\put(65,0){$\mbox{\boldmath$\delta=0.02$}$}\\[0.1cm]\hline
1~~&~~~~&~~~~&~~0.4900500 \\[0.05cm]
2~~&~~~~&~~~~&~~0.1152000 \\[0.05cm]
3~~&~~0.0460057~~&~~0.0460056~~&~~0.0460056 \\[0.05cm]
4~~&~~0.0220512~~&~~0.0220500~~&~~0.0220500 \\[0.05cm]
5~~&~~0.0112554~~&~~0.0112500~~&~~0.0112550 \\[0.05cm]\hline
\\
\put(65, 0){$\mbox{\boldmath$\delta=0.2$}$}\\[0.1cm]\hline
1~~&~~0.4049962~~&~~0.4050000~~&~~0.4050000 \\[0.05cm]
2~~&~~0.0450856~~&~~0.0450000~~&~~0.0450000 \\[0.05cm]
3~~&~~~~&~~~~&~~0.0005556 \\[0.05cm]
4~~&~~~~&~~~~&~~0.0112500 \\[0.05cm]
5~~&~~~~&~~~~&~~0.0450000 \\[0.05cm]\hline
\end{tabular}


\newpage
\begin{thebibliography}{99}
\bibitem{ref1}
F. Cooper, A. Khare, U. Sukhatme, Phys. Rep., \textbf{251} (1995) 267
\bibitem{ref2}
E. Schr\"{o}dinger, Proc. R. Irish Acad., \textbf{46A} (1940) 183
\bibitem{ref3}
L. Infeld, T. E. Hull, Rev. Mod. Phys., \textbf{23} (1951) 21
\bibitem{ref4}
V. B. Matveev, Lett. Math. Phys., \textbf{3} (1979) 217
\bibitem{ref5}
T. Imbo, U. Sukhatme, Phys. Rev. Lett. \textbf{54} (1985) 2184
\bibitem{ref6}
F. Cooper, P. Roy, Phys. Lett. A \textbf{143} (1990) 202
\bibitem{ref7}
A. Comtet, A. Bandrauk, D. K. Campbell, Phys. Lett. B \textbf{150} (1985)
159
\bibitem{ref8}
E. Kasap, B. G\"{o}n\"{u}l, M. \c{S}im\c{s}ek,
Chem. Phys. Lett., \textbf{172} (1990) 499
\bibitem{ref9}
E. D. Filho, R. M. Ricotta, Phys. Lett. A, \textbf{269} (2000) 269
\bibitem{ref10}
B. G\"{o}n\"{u}l, O. \"{O}zer, Y. Can\c{c}elik, M. Ko\c{c}ak,
Phys. Lett. A, \textbf{275} (2000) 238
\bibitem{ref11}
U. Laha, C. Bhattacharyya, K. Roy, B. Talukdar, Phys. Rev. C, \textbf{38},
(1988) 558
\bibitem{ref12}
L. E. Gendenstein, I. V. Krive, Usp. Fiz. Nauk \textbf{146} (1985) 553
\bibitem{ref13}
L. E. Gendenstein, JETP Lett. \textbf{38} (1983) 356
\bibitem{ref14}
F. Cooper, J. N. Ginocchio, A. Khare, Phys. Rev. D, \textbf{36} (1987)
2458
\bibitem{ref15}
R. Dutt, A. Khare, U. Sukhatme, Am. J. Phys. A, \textbf{56} (1988) 163
\bibitem{ref16}
G. Levai, J. Phys. A, \textbf{22} (1989) 267
\bibitem{ref17}
J. W. Dabrowska, A. Khare, U. Sukhatme, J. Phys. A, \textbf{21} (1998)
L195
\bibitem{ref18}
D. T. Barclay, R. Dutt, A. Gangopadhyaya, A. Khare, A. Pagnamenta,
U. Sukhatme, Phys. Rev. A, \textbf{48} (1993) 2786
\bibitem{ref19}
F. Cooper, A. Khare, J. Phys. A, \textbf{26} (1993) L901
\bibitem{ref20}
A. Khare, U. Sukhatme, J. Phys. A, \textbf{22} (1989) 2847
\bibitem{ref21}
B. Talukdar, U. Das, C. Bhattacharyaya, P. K. Bera , J. Phys. A,
\textbf{25} (1992) 4073
\bibitem{ref22}
C. V. Sukumar, J. Phys. A, \textbf{18} (1985) 2917
\bibitem{ref23}
S. Fl\"{u}gge, ``Practical Quantum Mechanics I'' (Springer-Verlag, New
York, (1974)
\bibitem{ref24}
H. E\u{g}rifes, D. Demirhan, F. B\"{u}y\"{u}kk{\i}l{\i}\c{c}, Physica
Scripta, \textbf{59}  (1999) 90
\bibitem{ref25}
H. E\u{g}rifes, D. Demirhan, F. B\"{u}y\"{u}kk{\i}l{\i}\c{c}, Physica
Scripta, \textbf{60} (1999) 195
\bibitem{ref26}
F. B\"{u}y\"{u}kk{\i}l{\i}\c{c}, H. E\u{g}rifes, D. Demirhan,
Theor. Chem. Acc., \textbf{98} (1997) 192
\bibitem{ref27}
A. F. Nikiforov, V. B. Uvarov, ``Special Functions of Mathematical
Physics'' (Birkhauser, Basel, 1988)
\bibitem{ref28}
D. A. Morales, Chem. Phys. Lett., \textbf{161} (1989) 253
\bibitem{ref29}
M. Bag, M. M. Panja, R. Dutt, Y. P. Varshni, \textbf{222}  (1994) 46
\bibitem{ref30}
A. Z. Tang, F. T. Chan Phys. Rev. A, \textbf{35} (1987) 911
\end{thebibliography}
\end{document}